\documentstyle[12pt]{article}
\begin{document}\title{$g/u(1)^d$ parafermions from constrained WZNW theories.}
\author{A.V.Bratchikov \thanks{bratchikov@elect.ru}\\
Kuban StateTechnological University,\\ 2 Moskovskaya Street,
Krasnodar, 350072,Russia. } \date{December,1997}
\maketitle \begin{abstract}
The conformal field theory
based on the $g/u(1)^d$ coset
construction is treated as the
WZNW theory for the affine Lie algebra $\hat g$ with
the constrained $\hat u(1)^d$ subalgebra.
Using a modification of the
generalized canonical quantization method generators and primary
fields of an extended symmetry algebra are found for arbitrary
d.\end{abstract} \section{Introduction.}

The $g/u(1)^d,1\le d\le rank\,g,$ parafermions are generators of an
exdended symmetry algebra of the conformal field theory based on the
$g/u(1)^d$ coset construction \cite{BH,GKO}
\begin{equation} K=L^g-L^{u(1)^d}.  \end{equation}
where $L^g$ is the energy-momentum tensor,
which is associated with the affine Lie algebra $\hat g$ via the Sugawara
formula. In the case of $su(2)/u(1)$ the
parafermions were constructed by Fateev and Zamolodchikov
\cite{FZ}.These authors found equations which relate the $su(2)/u(1)$
parafermions with the $\hat {su}(2)$ algebra.A generalization of these
equations to all simple Lie algebra was proposed in
ref.\cite{G}.However it was not proved that the generalized equations
define parafermions.

In a recent article \cite {B} we displayed the approach to the study of
the $g/u(1)^d$ conformal field theory which is based on the connection
between this theory and the Wess-Zumino-Novikov-Witten (WZNW) theory
for $\hat g$ with the constrained $\hat u(1)^d$ currents.It was shown
using an operator version of Dirac's procedure \cite {D} that the
initial energy-momentum tensor $L^g$ is replaced in the constrained
theory by $K.$ It turned out that the constrained $su(2)$ WZNW theory
is equivalent to the $su(2)/u(1)$ theory.However in the general case
the constrained currents determine the $g/u(1)^d$ parafermions only up
to some factors.

In the present paper we choose another way. We perform
operator quantization of the constrained WZNW theory using the approach
which is based on a modification of the generalized canonical
quantization method (see e.g.\cite {BFF} and references therein) and
show that the resulting symmetry algebra is an extended symmetry
algebra of the $g/u(1)^d$ coset conformal field theory.We present two
explicit realizations for the parafermions: a realization
where they are represented as elements of a coset algebra and a vertex
operator construction.Using the second realization we find the
equations which define the $g/u(1)^d$ parafermions through their
relationship with the generators of $\hat g$.The theory has a discrete
symmetry group which is generated by the $u(1)^r$ currents, where
$r=rank\,g$.The quantization method enables us to find primary fields
of the $g/u(1)^d$ theory.A representation of the parafermionic algebra
is obtained by applying the parafermions to the parafermionic primary
fields.

The parafermionic algebra is associated with a classical
quadratically nonlinear algebra which can be obtained using both the
generalized canonical quantization method and the Dirac-bracket
formalism.This demonstrates the equivalence of the quantization
methods.

The plan of the paper as follows.In Sec.2 we describe the quantization
method.Putting auxiliary operators into correspondence with
first and second class constraints we convert the initial
constraints into effective abelian constraints. The constrained
algebra is obtained by replacing the initial
operators by operators which commute with the effective abelian
constraints.In section 3 we review the WZNW theory and introduce the
constraints.In Sec.4 we perform quantization of the constrained WZNW
theory and find operators which replace the initial symmetry generators.
These operators are expressed in terms of the initial currents and free
bosonic fields.We show that the $g/u(1)^d$ parafermionic
algebra can be formulated as a coset algebra in a natural way and find the
parafermionic currents and primary fields.In Sec.6 we find vertex operator
construction for the $g/u(1)^d$ parafermionic algebra.In Sec.7 we consider
the constrained classical current algebra, find the corresponding Dirac
brackets and show that the same brackets can be obtained using the generalized
canonical quantization method.

\section{The quantization method.}

Let $T^{A}$ and
$T_{\mu}$ be first
and second class constraints.For our
purpose it is sufficient to consider constraints of zero Grassman
parity $\epsilon (T^{A})=\epsilon (T_{\mu})=0$ and assume
that they obey the commutator relations
\begin{eqnarray}
[T^{A},T^{B}]=[T^{A},T_{\mu}]=0, \qquad
[T_{\mu},T_{\nu}]=r_{{\mu}{\nu}},
\end{eqnarray}
where $r_{{\mu}{\nu}}$
is an inversible c-number matrix.

According to the prescription of the generalized canonical quantization
method \cite {BFF} we introduce the auxiliary operators $b_{\mu}$ which
obey the commutator relation \begin{equation}
[b_{\mu},b_{\nu}]=r_{{\mu}{\nu}} \end{equation} and convert the second
class constraints $T_{\mu}$ into the effective abelian constraints
\begin{equation}
\tilde T_{\mu}=T_{\mu}+ib_{\mu}.
\end{equation}
It is easy to see that
\begin{equation}
[\tilde T_{\mu},\tilde T_{\nu}]=0,\qquad
\tilde T_{\mu}(0)=T_{\mu}
\end{equation}
and
\begin{equation}
[\tilde T_{\mu},T^{A}]=0.
\end{equation}

Let us put into correspondence with each constraint $T^A$ a
pair of the canonical operators $(p^{A},q^{A})$ which, in contrast with
ref.  \cite{BFF}, are supposed to be of the same Grassman parity as the
constraints.In the extended phase space the operator $T^{A}$ can be
replaced by an operator $\tilde
T^{A}(\chi),\chi=(b_{\mu},p^{A},q^{A}),$ which, as well as $T^{A},$
satisfies the equations
\begin{equation} \label {gen.eq1}
 [\tilde T_{\mu},\tilde
T^{A}]=0,\qquad [\tilde T^{A},\tilde T^{B}]=0 \end{equation} and the
boundary conditions \begin{equation}
 \tilde
T^{A}(0)=T^{A}.\end{equation}

Let $F$ be an initial operator.We shall demand that $F$ is replaced in
the constrained theory by an operator ${\cal F}(\chi)$ which commutes
with the effective abelian constraints $(\tilde T^{A}, \tilde
T_{\mu}).$ The generating equations of the constrained algebra then
assume the form
\begin{eqnarray}\label {gen.eq}
[\tilde T_{\mu},{\cal F}]&=&0,\qquad
[\tilde T^{A},{\cal F}]=0,\nonumber \\
{\cal F}(0)&=&F.
\end{eqnarray}
Solutions of equations
(\ref {gen.eq1}) and (\ref {gen.eq}) are not completely fixed by the
boundary conditions.The arbitrariness is to be removed by a gauge
condition.

\section{
The WZNW model and constraints.}

The WZNW theory \cite{W,KZ} is
invariant with respect to two commuting affine Lie algebras, which are
generated by the left and right currents
$\left(H^s(z),E^\alpha(z)\right),$ $\left(\bar H^s(\bar z),\bar
E^\alpha(\bar z)\right)$,where $s=1,\ldots,r$ and $\alpha$
are roots of $g.$ Since the algebras generated by the left and right
currents are identical we shall only consider the first.The theory is
also invariant with respect to the conformal algebra.The Virasoro
generator $L^g(z)$ is expressed quadratically in terms of the currents.

The fields $H^s(z),E^\alpha(z)$ and $L^g(z)$
satisfy the operator product expansions
\begin{eqnarray}H^s(z)H^t(w)&=&{{k\delta^{st}}\over(z-w)^2},
\qquad
H^s(z)E^\alpha(w)={{\alpha^sE^\alpha(w)}\over{z-w}},\nonumber \\
E^\alpha(z)E^\beta(w)&=&\cases
{{{f(\alpha,\beta)
E^{\alpha+\beta}(w)}\over{z-w}},&if $\alpha+\beta$ is a root \cr
{k\over(z-w)^2}+{{\alpha\cdot{H(w})}\over{z-w}},&if
$\alpha+\beta=0$\cr 0, &otherwise\cr} \nonumber \\
L^g(z)H^s(w)&=&{H^s(w)\over
(z-w)^2}+{\partial_wH^s(w)\over{z-w}}, \nonumber \\
L^g(z)E^{\alpha}(w)&=&{E^{\alpha}(w)\over
(z-w)^2}+{\partial_wE^{\alpha}(w)\over{z-w}}, \nonumber \\
L^g(z)L^g(w)&=&{c_g\over
2(z-w)^4}+{2L^g(w)\over(z-w)^2}
+{{\partial_w L^g(w)\over{z-w}}}, \nonumber \\
c_g&=&{{kdim
g}\over{k+h^{\vee}}},
\end{eqnarray}
where k is the level of the representation,$\,f(\alpha,\beta)$ are the
structure constants and  $h^{\vee}$ is the dual Coxeter number of the
algebra $g$.

Let g be simply-laced.In this case the currents
$H^s(z),E^\alpha(z)$ and energy-momentum tensor $L^g(z)$ can be
expressed in terms of the bosonic fields
\begin{equation}\label{boson}
\varphi^{sj}(z)=x^{sj}-i{a^{sj}_o}logz+i\sum_{n\not =0}{a_{n}^{sj}\over
n}z^{-n},
\end{equation}
where $j=1,\ldots,k,\>$  and
\begin{equation}\label{gamma}
[x^{sj},a_{o}^{tl}]=i\delta^{st}\delta^{jl},\qquad
[a_{m}^{sj},a_{n}^{tl}]=m\delta^{st}\delta^{jl}\delta_{m+n,0}.
\end{equation}

The bosonic construction for $H^s(z),E^\alpha(z)$ and $L^g(z)$ read
\cite{FK,S,DHS}
\begin{eqnarray}
E^\alpha(z)&=&\sum^k_{j=1}:e^{i\alpha\cdot\varphi^j(z)}:c_{\alpha}^j,\qquad
H_s(z)=\sum^k_{j=1}i\partial_z\varphi^{sj}(z),\nonumber \\
L^g\left(z\right)&=&{1\over{2(k+h^{\vee})}}\biggl((1+h^{\vee})\sum_{s=1}^r
\sum_{j=1}^k:
(i\partial_z\varphi^{sj})^2:\\
&+&2\sum_{s=1}^r\sum_{i<j}^k:(i\partial_z\varphi^{sj})(i\partial_z\varphi^{sj}):
+2\sum_\alpha\sum_{i<j}^k:\exp(i\alpha\cdot(\varphi^i-
\varphi^j)):c^i_{\alpha}c^j_{-\alpha}\biggr),\nonumber
\end{eqnarray}
where :: denotes normal ordering with respect to the
modes of the bosons,$\alpha^2=2$ and $c_\alpha^j$ is a cocycle operator.

We shall consider the WZNW theory subject to the constraints
\begin{equation}H^{A}(z)\approx0,\end{equation}
where $A=1,\ldots,d.$
It is convenient to decompose the constraints into modes
\begin{eqnarray}
H^A(z)&=&H_{o}^Az^{-1}+\sum_{n\ne 0}{H_{n}^Az^{-n-1}},\nonumber \\
H_{o}^A&=&\sum^k_{j=1}{a^{Aj}_{o}},\qquad
H_{n}^A=\sum^k_{j=1}{a^{Aj}_{n}},
\end{eqnarray}
and consider an equivalent set of constraints:
\begin{equation}
H_{o}^A\approx0,\qquad H_{n}^A\approx0.
\end{equation}
The
operators $H_{o}^A,H_{n}^A$ obey the algebra
\begin{equation} \label{class}
[H_{o}^A,H_{o}^B]=[H_{o}^A,H_{n}^B]=0,\qquad
[H_{m}^A,H_{n}^B]=km\delta^{AB}\delta_{m+n,0}.
\end{equation}
It follows from these commutator relations that the
constraints $H_{o}^A$ are first class and $H_{n}^A$ are second
class.

\section{Quantization of the model.}
\subsection{Generating equations of the constrained symmetry algebra.}

To quantize the system we put into correspondence with the
constraints the operators $\chi=(b_{n}^A,p^A,q^A)$ which
satisfy the commutator relations \begin{equation}\label{gost}
[b_{m}^A,b_{n}^B]=km\delta^{{AB}}\delta_{m+n,0},\qquad
[q^A,p^B]=ik\delta^{AB} .
\end{equation}
The effective abelian constraints $\tilde H_{n}^A(b)$ are given by
\begin{equation}\label{ec1}
\tilde H_{n}^A=H_{n}^A+ib_{n}^A
\end{equation}
and we replace the first class constraints $H_{o}^A$ by the following
\begin{equation}\label{ec2}
\tilde H_{o}^A=H_{o}^A-ip^A.
\end{equation}

According to eq.(\ref{gen.eq}) the other operators of the constrained
symmetry algebra ${\cal H}^P(z)\equiv{\cal H}^P(\chi,z),$ where
$P=d+1,\ldots,r, {\cal E}^{\alpha}(z)\equiv{\cal E}^\alpha(\chi,z)$ and
${\cal L}^g(z)\equiv{\cal L}^g(\chi,z) $ satisfy the equations
\begin{eqnarray}
[\tilde H_{N}^A,{\cal H}^P]=0,\qquad
[\tilde H_{N}^A,{\cal E}^{\alpha}]=0,
\qquad
[\tilde H_{N}^A,{\cal L}^g]=0
\end{eqnarray} with the boundary conditions
\begin{equation}\label{c} {\cal H}^P(0,z)=H^P(z),\qquad {\cal
E}^{\alpha}(0,z)=E^\alpha(z),\qquad {\cal L}^g(0,z)=L^g(z).
\end{equation}

A solution of these
equations is given by
\begin{eqnarray} \label{concur} {\cal
H}^P(z)&=&H^P(z),\qquad
{\cal{E}}^{\alpha}(z)={E}^{\alpha}(z):e^{{1\over
k}\tilde\alpha\cdot\phi(z) }:,\nonumber \\
{\cal
L}^g(z)&=&K(z)+{1\over{2k}}\left(\tilde H^A(z)\right)^2,
\end{eqnarray}
where $\tilde \alpha=(\alpha^A),\,K$ is given by (1),
\begin{equation}\tilde H^A(z)=H^A(z)+\partial_z\phi^A(z)
\end{equation}
and $\phi=(\phi^A)$ are the bosonic fields:
\begin{equation}\phi^A(z)=q^A-ip^A\log{z}+i\sum_{n\not
=0}{{b_{n}^A\over n}z^{-n}}.  \end{equation} This solution is not the
general one.It can be fixed by appropriate gauge conditions.

\subsection{Parafermionic currents.}

The operator product expansions of $K$ with ${\cal H}^P$
and ${\cal E}^{\alpha}$ are
given by
\begin{eqnarray}\label{conf} K(z){\cal H}^P(w)&=&{{{\cal H}^P(w)}\over
(z-w)^2}+{{\partial_w{\cal H}^P(w)}\over{z-w}},\nonumber \\
K(z){\cal E}^{\alpha}(w)
&=&{{\Delta_\alpha{\cal E}^\alpha(w)}\over(z-w)^2}
+{1\over{z-w}}\left(\partial_w{\cal
E}^{\alpha}(w)-\frac 1 k\tilde\alpha\cdot\tilde H(w){\cal
E}^{\alpha}(w)\right), \end{eqnarray} where \begin{equation}
\Delta_\alpha=1-{{\tilde\alpha}^2\over{2k}}
\end{equation}

Let $\Omega$ be the algebra
gen\-er\-ated by ${\cal H}^P(z),{\cal E}^{\alpha}(z)$ and $K(z).$
Since these operators commute with $\tilde H^A_N$ so
therefore must all the fields of $\Omega$.
Let us define the following set of the fields
\begin{equation}\Upsilon
=\left(U\in\Omega;\,U\vert_{\tilde H^A=0}=0 \right).
\end{equation} One
can check that for arbitrary $U(z)\in \Upsilon$ and $X(z)\in \Omega$
\begin{equation}U(z)X(w)\in \Upsilon.\end{equation} Hence $\Upsilon$ is
an invariant subalgebra of $\Omega$ and the coset space
$\Omega/\Upsilon$ is an algebra.We shall denote by $\{X\}(z)$
the coset represented by the field $X(z).$

The coset field $\{K\}(z)$ satisfies the $g/u(1)^d$
Virasoro algebra:
\begin{equation}\{K\}(z)\{K\}(w)={c_{g/u(1)^d}\over
2(z-w)^4}+{2\{K\}(w)\over(z-w)^2}
+{{\partial_w \{K\}(w)\over{z-w}}},\end{equation}
where $c_{g/u(1)^d}=c_{g}-d.$
It follows from (\ref{concur}) that
$\{K\}(z)$ can be represented by ${\cal L}^g(z)$ as well.
Equations (\ref{conf}) express the fact that
$\{{\cal H}^P\}(z)$ and $\{{\cal E}^{\alpha}\}(z)$ are primary fields
of the $g/u(1)^d$ Virasoro algebra:
\begin{eqnarray}\{K\}(z)\{{\cal H}^P\}(w)&=&{\{{\cal H}^P\}(w)\over
(z-w)^2}+{{\partial_w\{{\cal H}^P\}(w)}\over{z-w}},\nonumber \\
\{K\}(z)\{{\cal E}^{\alpha}\}(w)
&=&{{\Delta_\alpha\{{\cal E}^\alpha\}(w)}\over(z-w)^2}
+\frac{\partial_w\{{\cal E}^{\alpha}\}(w)} {z-w},
\end{eqnarray}

The theory has the discrete symmetry group which is generated
by $H^s_o$:
\begin{eqnarray}
[H^s_o,\{{\cal H}^P\}]=0,\qquad [H^s_o,\{{\cal E}^\alpha\}]
=\alpha^s\{{\cal E}^\alpha\} \qquad [H^s_o,\{K\}]=0.
\end{eqnarray}

Let us consider the case of $su(2)/u(1)$.The $\hat {su}(2)$ generators
$E^+(z),E^-(z)$ and $H(z)$ are given by
\begin{eqnarray}
E^+(z)&=&\sum^k_{j=1}:e^{i\sqrt 2\varphi^j(z)}:,\qquad
E^-(z)=\sum^k_{j=1}:e^{-i\sqrt 2\varphi^j(z)}:,\nonumber \\
H(z)&=&\sum^k_{j=1}i\partial_z\varphi^{j}(z),
\end{eqnarray}
where $\varphi^j\equiv \varphi^{1j}.$
In the theory with the constrained current $H(z)\approx 0$ these
operators are replaced by
\begin{eqnarray}
{\cal{E}}^{+}(z)&=&{E}^{+}(z):e^{{\sqrt 2\over k}\phi(z) }:,\qquad
{\cal{E}}^{-}(z)={E}^{-}(z):e^{-{\sqrt 2\over k}\phi(z)
}:,\nonumber \\ \tilde H(z)&=&H(z)+\partial_z\phi(z).
\end{eqnarray}
We find that \begin{eqnarray}
{\cal{E}}^{+}(z){\cal{E}}^{-}(w)&=&k(z-w)^{-2+{2\over
k}}\Biggl(I+\tilde h(w)(z-w)\nonumber \\
&+&\biggl({{k+2}\over
k}K(w)+{1\over 2} \partial_w\tilde h(w)+{1\over
2}\tilde h^{2}(w)\biggr)(z-w)^2\Biggr), \end{eqnarray} where $\tilde
h=\tilde H /k$ and $K$ is given by the $su(2)/u(1)$ coset
construction. From this it follows that the coset currents
$\{{\cal{E}}^{+}\}(z)$ and $\{{\cal{E}}^{-}\}(z)$ satisfy the
parafermionic algebra of ref.\cite{FZ} \begin{equation}\label{}
\{{\cal{E}}^{+}\}(z)\{{\cal{E}}^{-}\}(w)=k(z-w)^{-2+{2\over
k}}\left(\{I\}+{{k+2}\over k}\{K\}(w)(z-w)^2\right).
\end{equation}
\subsection{Primary fields.}
Let  $G^{\Lambda}_{\lambda}(z)$ be the initial primary field
with highest weight $\Lambda$  which has
the weights $\lambda$
\begin{eqnarray}
E^{\alpha}_nG^{\Lambda}_{\lambda}&=&H^{s}_nG^{\Lambda}_{\lambda}=
L^g_nG^{\Lambda}_{\lambda}=0,\qquad n>0,\nonumber \\
H^{s}_oG^{\Lambda}_{\lambda}&=&\lambda^sG^{\Lambda}_{\lambda},\qquad
E^{\alpha}_oG^{\Lambda}_{\lambda}
=t^{\alpha}_{\lambda\lambda\prime}G^{\Lambda}_{\lambda\prime},\nonumber \\
L^g_oG^{\Lambda}_{\lambda}&=&{\Delta_\Lambda}G^{\Lambda}_{\lambda},\qquad
L^g_{-1}G^{\Lambda}_{\lambda}=\partial_{z}G^{\Lambda}_{\lambda},
\end{eqnarray}
where $t^{\alpha}$ are the step generators of $g$ in the representation
with highest weight $\Lambda$.The anomalous dimension $\Delta_\Lambda$
is given by \cite {KZ}
\begin{equation}
\Delta_\Lambda={\Lambda(\Lambda+\rho)\over{2(k+h^{\vee})}},
\end{equation}
where $\rho$ is half the sum positive roots of $g$.

Let us define
\begin{equation}{\cal G}^\Lambda_\lambda(z)
={G}^\Lambda_\lambda(z):e^{{{1\over k}
\tilde\lambda\cdot\phi(z)}}:,\end{equation}
where $\tilde\lambda=(\lambda^A).$
This field satisfies the
following equations \begin{eqnarray}\label{pri} \tilde H_{n}^A{\cal
G}^{\Lambda}_{\lambda}&=&0,\quad n\ge 0,\qquad
K_n{\cal G}^{\Lambda}_{\lambda}=0,\quad n>0,\nonumber \\
K_o{\cal G}^{\Lambda}_{\lambda}&=&{\Delta^\Lambda_\lambda}{\cal
G}^{\Lambda}_{\lambda},\qquad
K_{-1}{\cal
G}^{\Lambda}_{\lambda}=\partial_{z}{\cal
G}^{\Lambda}_{\lambda}-\frac 1 k
\tilde\lambda\cdot\tilde H_{-1}{\cal G}^{\Lambda}_{\lambda},
\end{eqnarray} where \begin{equation}\label{andim}
{\Delta^\Lambda_\lambda}=\Delta_{\Lambda}-
{{\tilde\lambda^2}\over{2k}}.
\end{equation}

Let $T$ be the space which is obtained by applying the
currents of $\Omega$ repeatedly to the fields
${\cal G}^\Lambda_\lambda.$ Let \begin{equation} V=\left({\cal G}\in
T;\,{\cal G}\vert_{\tilde{H}^A=0}=0\right).  \end{equation} The
space $V$ is an invariant subspace with respect to the algebra
$\Omega.$ Hence one can consider the representation of $\Omega$ in the
coset space $T/V.$ It follows from eq.(\ref {pri}) that the coset
$\{{\cal G}^\Lambda_\lambda\}$ which is represented by
${\cal G}^{\Lambda}_{\lambda}$ is the primary
field with the anomalous dimension
$\Delta^{\Lambda}_{\lambda}$ (\ref{andim}).

The space $T/V$ can be decomposed according to the transformation
properties under the $u(1)^r$ algebra.The field
$\{{\cal G}^{\Lambda}_{\lambda}\}$ belongs to the subspace with the
charge $\lambda$:
\begin{equation}
H_{o}^s\{{\cal G}^{\Lambda}_{\lambda}\}=\lambda^s\{{\cal
G}^{\Lambda}_{\lambda}\}.  \end{equation}

\section{Vertex operator construction for the parafermionic algebra.}

In this section we show that the theory for $g$ simply-laced can be
formulated in terms of the initial operators $x^{sj},{a^{sj}_o},$ and
$a_{n}^{sj}$.

The constraints (\ref{ec1}) and (\ref{ec2}) can be
considered as the strong operator equations
\begin{equation}\label{strong} p^A=\sum^k_{j=1}a_{o}^{Aj} \qquad
b_{n}^A=\sum^k_{j=1}a_{n}^{Aj}.  \end{equation} Substituting this into
(\ref{concur}) we express the constrained currents in terms of the
initial operators and the auxiliary operators $q=(q^A)$
\begin{eqnarray}\label{red} {\cal E}^\alpha(z)={\cal
E}_o^\alpha(z)e^{\frac{1}{k} \tilde\alpha\cdot q}, \end{eqnarray} where
\begin{eqnarray}\label{}
{\cal E}_o^\alpha(z)=\sum^k_{j=1}
:e^{i\left[\tilde\alpha\cdot\omega^j(z)+\tilde\alpha^{\bot}\cdot\varphi^j(z)
\right]}:c_{\alpha}^j,
\end{eqnarray}
\begin{eqnarray}\label{}
\omega^{Aj}(z)=x^{Aj}-i\eta^{jl}\left({a^{Al}_o}logz-\sum_{n\not
=0}{a_{n}^{Al}\over n}z^{-n}\right)
\end{eqnarray}
and
\begin{equation}\label{}
\eta^{ij}=\cases{{(k-1)\over k}&if $i=j,$\cr
{-{1\over k}} &if $i\not=j.$ \cr}
\end{equation}
Note that ${\cal E}^\alpha(z)$  (\ref{red}) belongs to the coset
$\{{\cal E}^\alpha\}(z)$.

Let us put into correspondence with
$\{K\},\{{\cal H}^P\}$ and $\{{\cal E}^\alpha \}$ the fields
${K},{H}^P$ and ${\cal E}^{\alpha}$ (\ref{red}) respectively.This map is the
isomorphism of the algebras.Since $q$ commute with all
the initial operators we can set $q=0$ in (\ref{red}).Therefore the
generators of $\Omega/\Upsilon$ can be represented by the
fields ${K},{H}^P$ and ${\cal E}_o^\alpha$ which are expressed
in terms of $x^{sj},{a^{sj}_o},$ and $a_{n}^{sj}.$ We have checked
that ${\cal E}_o^\alpha(z)$ satisfies the equation
\begin{eqnarray}
K(z){\cal E}_o^{\alpha}(w) ={{\Delta_\alpha{\cal
E}_o^\alpha(w)}\over(z-w)^2} +\frac{\partial_w{\cal
E}_o^{\alpha}(w)}{z-w}.
\end{eqnarray}

We find that the initial currents $E^\alpha(z)$ can be expressed
as follows
\begin{eqnarray}\label{km}
E^\alpha(z)=:e^{{i\over k}{
\tilde\alpha\cdot\tilde\varphi(z)}}:r^\alpha{\cal E}_o^\alpha(z),
\end{eqnarray} where
\begin{eqnarray}\label{}
r^\alpha=e^{-{i\over k}{\tilde\alpha\cdot x}},\qquad
x^A=\sum_{j=1}^kx^{Aj}
\end{eqnarray}
and
\begin{eqnarray}\label{}
\tilde\varphi^A=\sum_{j=1}^k\varphi^{Aj}
\end{eqnarray}
These equations can be used as the generating equations for the
$g/u(1)^d$ papafermions.

In the case of $su(2)$ the cocycles $c^j_\alpha$ can be replaced by
the identity operator.Since $x^1$ commutes with ${\cal E}_o^\alpha(z)$ we
can set $x^1=0$ and rewrite equations (\ref{km}) in the
form \cite{FZ} \begin{eqnarray}\label{} E^+(z)=\sqrt
k\psi^-(z):e^{i{\sqrt 2\over k}\tilde\varphi(z)}:,\nonumber\\
E^-(z)=\sqrt k\psi^+(z):e^{-i{\sqrt 2\over k}\tilde\varphi(z)}:,
\end{eqnarray} where \begin{equation}\psi^-(z)={1\over {\sqrt k}}{\cal
E}_o^+(z)\qquad \psi(z)= {1\over {\sqrt k}}{\cal E}_o^-(z)
\end{equation}
and $\tilde\varphi(z)\equiv\tilde\varphi^1(z).$

These results can be generalized to non-simply-laced algebras using
the vertex operator representation of the associated affine Lie algebras
\cite {GNOS,BT}.

\section{\bf Classical parafermions and the Dirac-bracket formalism.}

Quantization of the $g/u(1)^d$ theory can also be performed
by using the Dirac-bracket formalism.The approach presented
above is equivalent to the canonical quantization of the classical WZNW
theory.

To show this we first find a classical system
which corresponds to the $g/u(1)^d$ theory.Let us replace the operators
$H^s(z),E^\alpha(z)$ and $\phi^A(z)$ by the functions
$H^s(\sigma),E^\alpha(\sigma),\phi^A(\sigma)$, satisfying the Poisson
bracket relations \begin{eqnarray}\label{curr}
\{H^s(\sigma),H^t(\sigma^{\prime})\}&=&-k\delta^{st}\delta^\prime,
\qquad
\{H^s(\sigma),E^\alpha(\sigma^{\prime})\}=\alpha^sE^\alpha(\sigma)\delta,
\nonumber \\
\{E^\alpha(\sigma),E^\beta(\sigma^{\prime})\}&=&\cases {
f(\alpha,\beta)
E^{\alpha+\beta}(\sigma)\delta,&if ${\alpha+\beta}$ is a root \cr
-k\delta^\prime+\alpha\cdot{H(\sigma)}\delta,&if
$\alpha+\beta=0$\cr 0, &otherwise\cr},
\end{eqnarray}
\begin{eqnarray}\label{scal}
\{\phi^A(\sigma),\phi^B(\sigma^{\prime})\}
&=&{k\over 2}\delta^{AB}\epsilon,
\end{eqnarray} where
$\delta=\delta(\sigma-\sigma^{\prime})$ and $\epsilon=\epsilon(\sigma
-\sigma^{\prime}).$ From this it
follows that the constraints $H^A(\sigma)\approx 0$ are second class.The
abelian operator constraints $\tilde H^A(z)$ are replaced by $\tilde
H^A(\sigma)=H^A(\sigma)+\partial_\sigma\phi(\sigma)$ which satisfy
the relations
\begin{equation} \label{fcc} \{\tilde H^A(\sigma),\tilde
H^B(\sigma^{\prime})\}=0. \end{equation}

It it easy to check that the functions
\begin{eqnarray}\label{anz}
{\cal H}^P(\sigma)=H^P(\sigma), \qquad
{\cal{E}}^{\alpha}(\sigma) =E^\alpha(\sigma)e^{{1\over
k}\tilde\alpha\cdot\phi(\sigma)}
\end{eqnarray}
commute with $\tilde H^A(\sigma)$
\begin{equation}\label{sec}
\{\tilde H^A(\sigma),{\cal H}^P(\sigma^{\prime})\}=0,
\qquad \{\tilde H^A(\sigma),
{\cal E}^\alpha(\sigma^{\prime})\}=0.  \end{equation}

Using the Poisson bracket relations (\ref{curr}) and (\ref{scal}) we
find \begin{eqnarray} \label{qna} \{{\cal H}^P(\sigma),{\cal
H}^Q(\sigma^{\prime})\}&=&-k\delta^{PQ}\delta^\prime, \qquad
\{{\cal H}^P(\sigma),
{\cal E}^\alpha(\sigma^{\prime})\}=\alpha^P{\cal E}^\alpha(\sigma)\delta,
\nonumber \\ \{{\cal
E}^\alpha(\sigma),{\cal E}^\beta(\sigma^{\prime})\}&=&\cases {
f(\alpha,\beta) {\cal E}^{\alpha+\beta}(\sigma)\delta+ \cr +{\tilde
\alpha \cdot \tilde \beta \over{2k}}{\cal E}^\alpha(\sigma){\cal
E}^\beta(\sigma^{\prime})\epsilon, &if ${\alpha+\beta}$ is a root \cr
-k\delta^\prime+\tilde \alpha \cdot{\tilde H(\sigma)}\delta+\tilde
\alpha^\bot\cdot{\cal H}(\sigma)\delta- \cr -{{\tilde
\alpha}^2\over{2k}}{\cal E}^\alpha(\sigma){\cal
E}^{-\alpha}(\sigma^{\prime})\epsilon, &if $\alpha+\beta=0$ \cr
{\tilde \alpha \cdot
\tilde \beta\over{2k}}{\cal E}^\alpha(\sigma){\cal
E}^\beta(\sigma^{\prime})\epsilon,&otherwise, \cr}  \end{eqnarray}
where ${\cal H}(\sigma)=\left({\cal H}^P(\sigma)\right).$ Similar algebra
was found by the authors of
ref.\cite{BCR} using the gauged WZNW lagrangian.

The algebra $R$ defined by equations (\ref{fcc}),(\ref{sec})
and (\ref{qna}) can be obtained using the Dirac-bracket formalism. Let
$I$ and $J$ belong to the initial current algebra (\ref{curr}).In the
constrained theory they are replaced by the currents which satisfy the
Dirac bracket relations \begin{equation}\label{brack}
\{I,J\}_D=\{I,J\}-{\int{d\sigma
d\sigma^{\prime}\{I,H^A(\sigma)\}\{H^A(\sigma),H^B(\sigma^{\prime})\}^{-1}
\{H^B(\sigma^{\prime}),J\}}}.  \end{equation}
We find that  $H^s$ and $E^\alpha$ obey the algebra
\begin{eqnarray}\label{db}
\{H^A(\sigma),H^B(\sigma^{\prime})\}_D&=&
\{H^A(\sigma),H^P(\sigma^{\prime})\}_D=
\{H^A(\sigma),{E}^\alpha(\sigma^{\prime})\}_D=0,\nonumber \\
\{H^P(\sigma),H^Q(\sigma^{\prime})\}_D&=&-k\delta^{PQ}\delta^\prime,
\qquad
\{H^P(\sigma),E^\alpha(\sigma^{\prime})\}_D=\alpha^PE^\alpha(\sigma)\delta,
\nonumber \\
\{{E}^\alpha(\sigma),{E}^\beta(\sigma^{\prime})\}_D&=&\cases {
f(\alpha,\beta)
{E}^{\alpha+\beta}(\sigma)\delta+
\cr
+{\tilde \alpha \cdot
\tilde \beta \over{2k}}{E}^\alpha(\sigma){E}^\beta(\sigma^{\prime})\epsilon,
&if ${\alpha+\beta}$ is a root \cr
-k\delta^\prime+\alpha
\cdot{H(\sigma)}\delta-
\cr
-{{\tilde \alpha}^2\over{2k}}{E}^\alpha(\sigma){E}^{-\alpha}(\sigma^{\prime})
\epsilon,
&if $\alpha+\beta=0$ \cr
{\tilde \alpha \cdot
\tilde \beta\over{2k}}{E}^\alpha(\sigma){E}^\beta(\sigma^{\prime})\epsilon,
&otherwise \cr}
\end{eqnarray}
This algebra is isomorphic to $R.$ The corresponding map is
\begin{eqnarray}\label{}
H^A(\sigma)\to \tilde H^A(\sigma),\qquad H^P(\sigma)\to
{\cal H}^P(\sigma),\qquad E^\alpha(\sigma)\to {\cal E}^\alpha(\sigma).
\end{eqnarray}

To quantize the system we could therefore start
from the Dirac brackets (\ref{db}) (or equivalently from the Poisson
brackets (\ref{fcc}),(\ref{sec})
and (\ref{qna})), use anzatz (\ref{anz}) and replace
the functions $H^s(\sigma),E^\alpha(\sigma)$ and $\phi^A(\sigma)$
by the corresponding operators.

\section{
Conclusion.}

Treating the $g/u(1)^d$ conformal field
theory as the WZNW theory for $\hat g$ with the constrained
$\hat u(1)^d$ currents and using a modification of the generalized
canonical quantization method we have found the explicit construction
of the symmetry generators of the $g/u(1)^d$ theory.The presented
quantization may be generalized to the WZNW theory for $\hat g$
with an arbitrary constrained current algebra $\hat h\subset \hat g.$
We expect that in the general case the conformal invariant solution of
the generating equations determines a symmetry algebra of the
$g/h$ conformal field theory.

\end{document}